\documentstyle [twoside,12pt]{article}
\newcommand{\nc}{\newcommand}
\nc{\la}{\lambda} \nc{\La}{\Lambda}  \nc{\al}{\alpha}
\nc{\te}{\theta}  \nc{\be}{\beta}	\nc{\ka}{\kappa}
\nc{\ga}{\gamma}  \nc{\Ga}{\Gamma}
\nc{\de}{\delta}  \nc{\De}{\Delta}
\nc{\si}{\sigma}  \nc{\Si}{\Sigma}
\nc{\om}{\omega}  \nc{\Om}{\Omega}
\nc{\nf}{\infty}   \nc{\nl}{\newline}
\nc{\ra}{\longrightarrow}
\nc{\beq}{\begin{equation}}
\nc{\eeq}{\end{equation}}
\nc{\beqa}{\begin{eqnarray}}  \nc{\dst}{\displaystyle}
\nc{\eeqa}{\end{eqnarray}} \nc{\nnb}{\nonumber}
\title{\bf Regularisation : many recipes,\\ but a unique principle :\\ Ward
identities and Normalisation conditions. \\ The case of CPT violation in
QED.}
\author{Guy Bonneau\thanks {\noindent Laboratoire de Physique Th\'eorique et
des Hautes Energies,
 Unit\'e associ\'ee au CNRS UMR 7589, Universit\'e Paris 7,
 2 Place Jussieu, 75251 Paris Cedex 05. Tel: 33-144277957,  Fax: 33-144277990, Email: bonneau@lpthe.jussieu.fr}}
\begin{document}
\date{}
\maketitle
\begin{abstract}
\noindent We analyse the recent controversy on a possible Chern-Simons like
term generated through radiative corrections in QED with a CPT  violating
term : we prove that, if the theory is correctly defined  through Ward
identities and normalisation conditions, no Chern-Simons term appears,
without any ambiguity. This is related to the fact that such a term is a kind of minor modification of the gauge fixing term, and then no  renormalised. The past year literature on that subject is
discussed, and we insist on the fact that any absence of an {\sl a priori}
divergence should be explained by some symmetry or some non-renormalisation
theorem.
\end{abstract}



\vfill {\bf PAR/LPTHE/00-08}\hfill  February 2000
\newpage

\section{Introduction}

In the past years, the interesting issue of a possible spontaneous breaking
of Lorentz invariance at low energy has been considered : this issue also led to CPT
breaking \cite{{CFJ90},{ColKos},{ColGlas}}. In particular, the general
Lorentz-violating extension of the minimal
$SU(3)\times SU(2)\times U(1)$ standard model has been discussed : as many
breaking terms are allowed, people look for possible constraints coming from
experimental results as well as from renormalisability requirements and
anomaly cancellation.

In that respect, there arose a controversy on a possible Chern-Simons like
term generated through radiative corrections
\cite{{JK99},{ColKos},{ColGlas},{Chen},{ChungOh},{Chung-Perez}}. This
phenomenon was studied in QED, an abelian gauge theory, as a part of the
standard model. We think that the controversy comes from some
misunderstandings on regularisation and (re)normalisation of a given theory
(see a review in \cite{GB1990}), and this paper intends to clarify the
situation.

In the next Section, the basis facts are recalled, with emphasis on Ward
identities, and the distinct three concepts of counterterms (finite or not),
quantum corrections or spurious anomalies and physical radiative corrections
are defined. In Section 3, breaking terms are added to the usual QED
Lagrangian density, and one loop contributions are discussed. We check that,
as soon as the theory is precisely defined through its symmetries
({ Ward identities}) and its physical parameters
({ Normalisation conditions}), there remains no ambiguity in the
results. Then we prove a kind of non-renormalisation theorem that allows for
a ``minimally CPT broken" theory (Subsect.3.2). Finaly, the literature is
discussed and some comments are also offered on the invoked difference
between an invariance of the action that leaves the Lagrangian density
non-invariant, and one that leaves the Lagrangian density invariant.

\section{Quantum electrodynamics} The Lagrangian density is~:
\beq\label{QED1} {\cal L}_0 = \bar{\psi}(i\not{\partial} - m - e\not{A})\psi
-
\frac{1}{4}F_{\mu\nu}^2 -\frac{1}{2\al}(\partial A)^2 +\frac{1}{2}\la^2
A_{\mu}^2\ ,\eeq where $\al$ is the gauge parameter and $\la$ an infra-red
regulator photon mass. The classical field equations write~:
\beq\label{eqfermion} (i\stackrel{\rightarrow}{\not\partial} - m)\psi(x) =
e\not{A}(x)\psi(x)\,,\quad
\bar{\psi}(x)(-i\stackrel{\leftarrow}{\not\partial} - m) =
e\bar{\psi}(x)\not{A}(x)\,,
\eeq
\beq\label{eqphoton} [\Box + \la^2]A_{\mu}(x)  -(1
-\frac{1}{\al})\partial_{\mu}(\partial_{\nu}A^{\nu}(x)) =
e\bar{\psi}(x)\ga_{\mu}\psi(x)\,,
\eeq  and the Feynman rules are~:
\beqa\label{Feynman} <A_{\mu}(-p)A_{\nu}(p)>_{class.}& = &
D^0_{\mu\nu}(p,-p)_{class.} =
\frac{-i}{p^2 - \la^2}(g_{\mu\nu} - \frac{p_{\mu}p_{\nu}}{ p^2 }) +
\frac{-i\al}{ p^2 -  \al\la^2}\frac{p_{\mu}p_{\nu}}{p^2}\nnb \\
<\psi(-p)\bar{\psi}(p)>_{class.} & = & S_0 (p,-p)_{class.} =
\frac{i}{\not p - m}\\
<\bar{\psi}(p)\psi(q)A_{\rho}(-(p+q))>_{class.}^{prop.} & = &
\Ga^{\rho}(p,\,q,\,-(p+q))_{class.}  = -ie\ga_{\rho}\,.\nnb \eeqa The Ward
identity resulting from gauge invariance ensures that the non-transverse
part of the 2-photon proper Green function
$\Ga_{\mu\nu}(p,-p)^{class.} =  -i(g_{\mu\nu}p^2 - p_{\mu}p_{\nu}) +i\la^2
g_{\mu\nu}  -\frac{i}{\al}p_{\mu}p_{\nu}$ is not renormalised, {\sl i.e.} the
parameters
$\al$ and $\la^2$ are ( unphysical) tree level parameters.

 By power counting, the primitively divergent graphs are here
$\Ga_{\mu\nu}(p,-p)_{transverse}\,,\
\Sigma(p,-p)$ and $\Ga^{\rho}(p,q,-(p+q))\,,$ respectively the transverse
photon and electron 2-points proper Green functions and the photon-electron
proper vertex function.  All corresponding parameters (positions and
residues of the poles in propagators, couplings at zero momenta,..) - but
for the unphysical, non renormalised ones
$\al$ and
$\la^2$ - require normalisation conditions, a point which has often been
missed since the successes of {  minimal} dimensionnal regularisation
scheme
\cite{GB1990}.  In perturbative quantum expansion, this requires addition of
{ definite counterterms into the Lagrangian density} : the question of
their being finite or not being merely a question of personal taste. For
example, in the original BPHZ \cite{BPHZ} substraction scheme, a definite
Taylor expansion with respect to external momenta  is subtracted from the
Feynman integrand so that the integration over loops momenta becomes
possible : some finite counterterms are then needed to implement the wanted
normalisation conditions (of course, they will depend on some
renormalisation scale
$\mu$ used to fix the normalisation conditions). In other schemes, infinite
counterterms (plus finite ones of course !) are defined after some
regularisation of the divergent integrals in order to achieve the same aim.
In the BRS approach, the number of primitive divergences and corresponding
counterterms is given by the dimension of the Fadeev-Popov 0-charge sector of
the cohomology space of the Slavnov operator corresponding to the isometries
that define the theory
\cite{Becchi}.

The second problem is related to the symmetries of the classical action : the
physical meaningfulness of the quantum extension requires that the symmetries
still hold at the quantum level, which is possible as soon as the
coresponding Ward identity has no { anomaly}. As is well known, this may
involve the addition of finite counterterms to the Lagrangian density ( the
so-called { quantum  corrections} or { spurious anomalies}). In
particular, each time one regularises a theory without respecting its
symmetries, such non-symmetric quantum corrections are needed, and moreover
the classical currents have to be redefined (renormalised) in order that
their conservation should lead to the correct Ward Identity (the correct
"contact terms"......). A pedagogical example may be found in the second
reference in
\cite{GB1981}. In the BRS approach, the absence of anomaly corresponds to an
empty Fadeev-Popov charge-1 sector of the cohomology space of the upper
mentioned Slavnov operator.

Finally, the success of a perturbative theory lies in its ability to compute
with precision some quantities such as S matrix elements whose classical
values acquire { radiative corrections}, for example  the
Bremsstrahl\"ung in annihilation processes :
$e^+\ +\ e^-\ \rightarrow \ {\rm photon}\ + {\rm final\ state}\ X\,,$ or the
magnetic moment of the electron which is found to be 2 in the classical
theory with Lagrangian (\ref{QED1}) and gets some definite corrections in
higher orders in the electric charge, or the Lamb-shift that results from
higher loop contributions to the photon self-energy $\Ga_{\mu\nu}\,,$
e.t.c...

\vspace{3mm}

One  should not confuse these three concepts : Lagrangian counterterms
(finite or not) required to get a definite perturbative expansion with
definite values for the physical parameters of the theory ; Lagrangian
finite spurious anomalies or quantum corrections required to compensate for
the use of a non symmetry preserving regularisation scheme ; calculable
radiative corrections to physical processes.

For instance, in QED the electromagnetic current is conserved, thanks to
the gauge invariance of the theory ; so one obtains the Ward identity :
\beqa\label{WI2}
p^{\mu}\Ga_{\mu\,;..\nu_i..}^{L+1;\,2N}(p,\,p_i\,;\,q_j\,;q'_j\,) & = &
-\frac{i}{\al}(p^2 - \al\la^2)p_{1\,\nu_1}\de_L^1\de_N^0 +\\ +
e\sum_{k=1}^{N} [\Ga_{...\nu_i...}^{L;\,2N}(p_i\,;..\,q_j...\,,q_k
+p\,,..\,;q'_j\,) &  - &
\Ga_{..\nu_i...}^{L;\,2N}(p_i\,;\,q_j\,;..q'_j..\,,q'_k +p\,,..)] \nnb
\eeqa where $L$ is the number of external photons (
 of momenta $p_i\,:\,i = 1,2,..L)$ and $N$ the number of incoming and
outgoing electrons ( of momenta $q'_j\,:\,j = 1,2,..N)\,,$ all momenta being
ingoing ones. In particular, for $L=1\,,\ N=0$ this gives the announced
non-renormalisation of the longitudinal part of the photon propagator
$-\frac{i}{\al}(p^2 -
\al\la^2)\,.$ This Ward identity is best rewritten on $\Ga\,,$ the generating
functional for proper Green functions~:
\beq\label{WI1}
\partial_{\mu}^x\frac{\de\Ga}{\de A_{\mu}(x)} - i\,e[\bar{\psi}_{\al}(x)
\frac{\de\Ga}{\de \bar{\psi}_{\al}(x)} - \psi_{\be}(x)
\frac{\de\Ga}{\de \psi_{\be}(x)}] = \frac{1}{\al}[\Box +
\al\la^2](\partial_{\mu}A^{\mu})\,.
\eeq Then, if one uses the BPHZ scheme that breaks gauge invariance, the
addition of finite counterterms into the Lagrangian and a redefinition of the
electromagnetic current is  required in order for the Ward Identity (\ref{WI1})
to be satisfied to all orders of perturbation theory. On the contrary,
Pauli-Villars or dimensional regularisation only need symmetric counterterms
(the usual Z factors).

In the same spirit, a softly broken axial symmetry exists at the classical
level ; but it does not survive quantisation because of the axial anomaly.
This one is readily seen, for example within dimensional regularisation
\cite{{DR1},{GB1981}}, to result from the impossibility of defining in an
algebraically consistent way (in { complex} D dimensions) a matrix
$\ga^5$ that anticommutes with all Dirac matrices $\ga^{\mu}\ :
\ga^{\mu}\ga_{\mu} = D$ \cite{BM}. So, when computing
$$\partial_{\mu}[\bar{\psi}\ga^{\mu}\ga^5 \psi](x)$$ with the help of the
electron field equations, an ``evanescent" contribution
\cite{{DR1},{GB1981}} appears
 :
$$[\bar{\psi}\hat{\ga}^{\mu}\stackrel{\leftrightarrow}{D_{\mu}}\psi](x)$$
where
$2\hat{\ga}^{\mu} = \{\ga^{\mu}\,,\,\ga^5\}$ and $D_{\mu}$ denotes the
covariant derivative $\partial_{\mu} + i\,e\,A_{\mu}\,.$ Moreover the axial
current has also to { be redefined}, a point often missed  (see for
example controversies on Adler Bardeen theorem in super-Yang-Mills between
1982 and 1985 \cite{GB1990} and a recent paper by Jacquot
\cite{Jacquot}).

\section{QED with odd-CPT Lorentz violating terms} As explained in the
introduction Section, let us now consider QED (equ.\ref{QED1}) with possible
presence of CPT-odd, Lorentz violating terms :
\beqa\label{viol1} {\cal L}_1(x) & = & -
b^{\mu}\bar{\psi}(x)\ga_{\mu}\ga^5\psi(x)\,,\quad\quad {\rm where}\  b^{\mu}\
{\rm is\ a\ fixed\ vector,}\nnb\\ {\cal L}_2(x) & = &
\frac{1}{2}c^{\mu}\epsilon_{\mu\nu\rho\si}
F^{\nu\rho}(x)A^{\si}(x)\,,\quad\quad {\rm where}\  c^{\mu}\ {\rm is\ a\
fixed\ vector.}\eeqa Other breakings could be considered (see a discussion
in the first paper of \cite{ColKos}), but we simplify and require charge
conjugation invariance, which selects
${\cal L}_1(x)$ and ${\cal L}_2(x)\,.$ Note for further reference that
experiments on the absence of birefringence of light in vacuum put very
restrictive limits on the value of $c^{\mu}\,,$ typically for a timelike
$c^{\mu}\,,\ c^{\small 0}/m \le 10^{-38 }$ \cite{ColKos}.

Let us consider the classical Lagrangian density ${\cal L}_0 + {\cal L}_1
+ {\cal L}_2\,.$ In order to avoid the difficulties resulting from the new
poles in the propagators, we take into account the smallness of the
breakings and include them into the { interaction Lagrangian density} as
super-renormalisable couplings. Moreover, we define the photon and electron
masses by the same normalisation conditions as in ordinary Q.E.D., $e.g.$
$$\dst <\psi(p)\bar{\psi}(-p)>^{prop.}\mid_{\not{p} = m\,,\;b = c  =0}\quad
=\ 0\,,\quad\cdots$$
According to standard results in renormalisation theory, these breakings add
new terms in the primitively divergent functions and require 2 new
normalisation conditions to fix their parameters
$b^{\mu}$ and
$c^{\nu}\,:$
\beqa\label{nor1} b^{\mu} &  = & - \frac{i}{4}
Tr[\ga^{\mu}\ga^5<\psi(p)\bar{\psi}(-p)>^{prop.}]\mid_{p=0}\,,\nnb \\
  c^{\mu} & = &
\frac{1}{12}\epsilon^{\mu\nu\rho\si}\frac{\partial}{\partial p^{\si}}
<A_{\nu}(p)A_{\rho}(-p)>^{prop.}\mid_{p=0}\,.\eeqa

\noindent Note that, contrary to  ${\cal L}_1(x)\,,$ the ${\cal L}_2(x)$ term
also breaks the local gauge invariance of the Lagrangian density, even if the
action remains gauge invariant. If fields and the gauge parameter function
$\La(x)$ vanish sufficiently rapidly at infinity, there will be no
difference ; however, the literature on this subject \cite{JK99} emphasizes
the difference, and we want to clarify this point. We shall prove that the
variation of ${\cal L}_2(x)$ in a local gauge transformation being linear in
the quantum field, no essential difference occurs.

So it is tempting to separate the discussion into two cases :
\begin{itemize}
\item QED with the two CPT odd, Lorentz breaking, C conserving terms :
${\cal L}_1(x)$ and
${\cal L}_2(x)\,.$
\item  QED with the sole breaking term
${\cal L}_1(x)$ : is it a consistent quantum theory ?
\end{itemize}

\subsection{Lagrangian of QED with CPT-odd, Lorentz and gauge breaking
terms.} Starting from the Lagrangian density  ${\cal L}_0\ + {\cal L}_1\ +
{\cal L}_2\,,$  let us analyse the Ward identity corresponding to the gauge
invariance of the action. The classical field equations are now~:
\beq\label{eqmodfermion} (i\stackrel{\rightarrow}{\not\partial} - m
-\not{b}\ga^5)\psi(x) = e\not{A}(x)\psi(x)\,,\quad
\bar{\psi}(x)(-i\stackrel{\leftarrow}{\not\partial} - m -\not{b}\ga^5) =
e\bar{\psi}(x)\not{A}(x)\,,
\eeq
\beq\label{eqmodphoton} [\Box + \la^2]A_{\mu}(x)  -(1
-\frac{1}{\al})\partial_{\mu}(\partial_{\nu}A^{\nu}(x)) -
\epsilon_{\mu\al\be\de}c^{\al}F^{\be\de}(x) =
e\bar{\psi}(x)\ga_{\mu}\psi(x)\,;
\eeq  then equations (\ref{eqmodfermion}) ensure that the vector current
$[\bar{\psi}\ga_{\mu}\psi](x)$ is conserved, and equation (\ref{eqmodphoton})
the fact that the ``scalar" field
$\partial_{\nu}A^{\nu}(x)$ is a free field of squared mass
$\al\la^2$ .

After addition of adequate counterterms, these equations of motion may
always be extended to the quantum level
\cite{{BPHZ},{BM}}, and, as a consequence, the same is true for the vector
current conservation and the free-character of the longitudinal photon
$\partial_{\nu}A^{\nu}(x)\,.$

Let $\Ga$ be the classical action
$$\dst\Ga = \int d^4x [{\cal L}_0\ + {\cal L}_1\ + {\cal L}_2]\,,$$ the Ward
identity  writes :
\beqa\label{WI4}
\lefteqn{\int d^4 x
\left\{\frac{1}{e}\partial_{\mu}\La(x)\frac{\de\Ga}{\de A_{\mu}(x)}+
i\La(x)[\bar{\psi}(x)\frac{\stackrel{\rightarrow}{\de}\Ga}
{\de\bar{\psi}(x)} -
\frac{\Ga\stackrel{\leftarrow}{\de}}{\de{\psi}(x)}\psi(x)]\right\} =}
\nnb
\\  & =& \int d^4 x \left\{-\frac{1}{e\al}
\partial_{\mu}A^{\mu}(x)\Box\La(x)  +
\frac{\la^2}{e}A^{\mu}(x)\partial_{\mu}\La(x) +
\frac{1}{2e}\epsilon_{\al\be\de\mu}c^{\al}F^{\be\de}(x)
\partial^{\mu}\La(x)\right\}
\nnb
\\ & \Rightarrow & W_x\,\Ga\, \equiv \partial_{\mu}\frac{\de\Ga}{\de
A_{\mu}(x)} - ie[\bar{\psi}(x)\frac{\stackrel{\rightarrow}{\de}\Ga}
{\de\bar{\psi}(x)} -
\frac{\Ga\stackrel{\leftarrow}{\de}}{\de{\psi}(x)}\psi(x)] =
\frac{1}{\al} [\Box +
\al\la^2]\partial_{\mu}A^{\mu}(x)\,.
\eeqa The last equation is exactly the same as the one for ordinary QED
(\ref{WI1}) : { so, to select the desired action, one needs an extra
symmetry} such as Lorentz invariance if one wants ordinary  QED (\ref{QED1}),
{ or some non-renormalisation theorem} if one wants to consistently
suppress the
${\cal L}_2$ term through a normalisation condition (Subsec. 3.2).

As soon as we use a regularisation that respects the symmetries (gauge,
Lorentz covariance and charge conjugation invariance), the perturbative
proof of renormalisability reduces to the check that the
${\cal{O}}(\hbar)$ quantum corrections to the classical action $\Ga\ :
\Ga_1 = \Ga|_{class.} + \hbar\De\,,$ constrained by the Ward identity
(\ref{WI4})~:
\beq\label{slavnov}
  W_x\,\De\, \equiv \partial_{\mu}\frac{\de\De}{\de A_{\mu}(x)} -
ie[\bar{\psi}(x)\frac{\stackrel{\rightarrow}{\de}\De} {\de\bar{\psi}(x)} -
\frac{\De\stackrel{\leftarrow}{\de}}{\de{\psi}(x)}\psi(x)] = 0\,,\eeq may be
reabsorbed into the classical action through suitable renormalisations of the
fields and parameters of the theory . Thanks to the quantum action principle
\cite{{Lowenstein},{Becchi}}, $\De$ is  a charge conjugation invariant
integrated local polynomial in the fields, their derivatives and the
parameters of the theory, of dimension 4 (recall that the photon field and
the parameters
$m,\ b^{\mu}\,,\ c^{\nu}$ have dimension 1, the electron field dimension
3/2) : then the general solution of (\ref{slavnov}) is readily shown to be
of the same form as the classical action $\Ga$ (without the gauge fixing and
photon mass terms) \hfill$ Q.E.D.$

The breakings (\ref{viol1}) introduce two operators which may be
defined through a modification of the classical action : to
$\Ga$
 we add two source terms  for the C = +1, dimension 3
insertions
$$ J^5_{\mu} = [\bar{\psi}\ga_{\mu}\ga^5\psi](x)\ {\rm
and}\
K^5_{\mu} =
[\frac{1}{2}\epsilon_{\mu\nu\rho\si}F^{\nu\rho}A^{\si}](x)\,:\quad
\dst\tilde{\Ga} = \Ga + \int d^4x \,\left[
\al^{\mu}(x)J^5_{\mu}(x) +
\be^{\mu}(x)K^5_{\mu}(x)\right]\,.$$

\noindent  Then, as soon as the
renormalisation is properly done, the operators being { bilinear in the
quantum fields}, the Ward identity (\ref{WI4}) holds true, to all orders of
perturbation theory (all orders in
$\hbar\,,\ e\,,\ b^{\mu}\,,\ c^{\nu}\,,\ m$), for the Green
functions with one insertion of either of these operators :
\begin{itemize}\item action on (\ref{WI4}) of
$\dst\frac{\de}{\de
\al^{\la}(y)}$ for the gauge invariant  axial current
$J^5_{\mu}(y)\,,$ the right-hand side
vanishing : $\dst W_x \frac{\de\Ga}{\de\al^{\la}(y)}\,\mid_{\al = \be =
0}\quad = 0\,,$
\item action on (\ref{WI4}) of
$\dst\frac{\de}{\de\be^{\la}(y)}$ for the non gauge invariant  operator
$K^5_{\la}(y)\,,$ the right-hand side reducing itself to a tree
contribution as the
variation is linear in the photon field :
$\dst W_x \frac{\de\Ga}{\de\be^{\la}(y)}\,\mid_{\al = \be = 0}\quad =
-\frac{1}{2}\epsilon_{\la\nu\rho\si}F^{\nu\rho}(y)
\partial_y^{\si}\de(y-x)\,.$
\end{itemize}
All Green functions are of
course computed with the { complete} Lagrangian density  ${\cal L}_0\ +
{\cal L}_1\ + {\cal L}_2\,.$ Of course, as for
ordinary QED, the axial current is not coupled to the fields of the model,
and its quantum non-conservation due to the axial anomaly is not dangerous.
However, should one consider the CPT breaking extensions of the standard
$SU(3)\times SU(2)\times U(1)$ model, the generalisation of the Adler-Bardeen
non renormalisation theorem would be necessary in order for the one loop
cancellation of the axial anomaly to stay to all orders.

As particular consequences of the previous discussion, the complete proper
2-photon Green function
$\Ga_{\mu\nu}(p,\,-p)$ satisfies
\beq\label{WI3a} p^{\mu}\Ga_{\mu\nu}(p,\,-p) = p^{\nu}\Ga_{\mu\nu}(p,\,-p)
=0\,,\quad\quad{\rm up\ to\ the\ classical\ longitudinal\
contribution\,,}\eeq
the one with one
axial-current insertion verifies :
\beqa\label{WI3}
p^{\mu}<A_{\mu}(p)A_{\nu}(q)[\bar{\psi} \ga^{\rho}\ga^5\psi](-(p+q))>^{prop.}
& = & \nnb\\
q^{\nu}<A_{\mu}(p)A_{\nu}(q)[\bar{\psi}
\ga^{\rho}\ga^5\psi](-(p+q))>^{prop.}& = &0\,,
\eeqa and the one with a $K^5_{\mu}$ insertion
\beqa\label{WI3bis}
p^{\mu'}<A_{\mu'}(p)A_{\mu}(q)[\frac{1}{2}\epsilon_{\rho\nu\si\la}
F^{\nu\si}A^{\la}](-(p+q))>^{prop.}
& = & \nnb \\
 q^{\mu'}<A_{\mu}(p)A_{\mu'}(q)[\frac{1}{2}\epsilon_{\rho\nu\si\la}
F^{\nu\si}A^{\la}](-(p+q))>^{prop.}& = &
- i\epsilon_{\mu\rho\si\la}p^{\si}q^{\la}\,.
\eeqa Note that { the validity of these
equations is not a question of personal taste or choice} : if the
renormalisation is correctly done, they do hold true as a
consequence of the gauge invariance of the action (moreover, as we shall
show, in the absence of (\ref{WI3},\ref{WI3bis}), the finiteness of the
CPT-odd part of the photon self energy will remain accidental).

\vspace{5mm}
We now illustrate these results by a one-loop calculation.
We first need a regularisation to give a meaning to the loops integrals
involved in those Green functions. Any consistent one is as good as any
other, but it is simpler to consider regularisations that respect the
invariances of the classical theory, here gauge invariance. A question
arises about Lorentz non-invariance : {\sl a priori} there is no longer
symmetric integration and averaging formulas for
$l^{\mu}l^{\nu}$ (where $l$  is a loop momentum variable) : any computation
will become fairly hard ; in particular, linear divergences would stay
\footnote{\ For instance as :
$$\dst\int_{-\La}^{\La'}\frac{x\,dx}{\sqrt{(x+p)^2 + m^2}} =  \La' -
\La + 2p - p\log{\frac{4\La\La'}{m^2}} + {\cal O}(1/\La,\,1/\La')\,.$$} and
require extra subtractions and counterterms.  However, as discussed in
\cite{ColKos}, the spirit of the spontaneously broken Lorentz invariance is
that, except for the vacuum expectation values $b^{\mu}$  and $c^{\nu}$ of
some fields, Lorentz invariance holds : Colladay and Kostelecky speak of true
``observer Lorentz invariance". So we shall use the Lorentz preserving
consistent dimensional regularisation of t'Hooft and Veltmann
\footnote{\, Let us emphasize that if one uses correct (anti-)commutation
relation of Dirac matrices $\ga^{\mu}$ and
$\ga^5$(Breitenlhoner and Maison
\cite{BM}), there will be no ambiguity in the results and a minimal
subtraction can be safely done (it corresponds to some implicit but definite
normalisation conditions).}
\cite{{DR1},{BM},{GB1981},{GB1990}} and, in particular, check at the one-loop
order that the correction to the self-energy of the photon of first order in
$b^{\mu}$  unambiguously vanishes at $p^2 = 0\,,$ in agreement with the
theorem in the appendix of
\cite{ColGlas}.

Note that gauge invariance (\ref{WI3a}) ensures, as in
standard Q.E.D., that the 4-photon Green function is not primitively
divergent . Then $\hbar\De{\cal L}_0$, the standard counterterms of
Q.E.D.,
 should be added to
${\cal L}_0\,.$ Let us now compute the
$b^{\mu}$ and
$c^{\nu}$ dependent counterterms, to one-loop order.

\subsubsection{The one-loop electron self-energy}
To first  order in the small breaking parameters
$b^{\mu}$ and $c^{\mu}\,,$ a one-loop calculation gives :
\beqa\label{propfermion}
\lefteqn{ -i<{\psi}(p)\bar{\psi}(-p)>^{prop.} =  [\not{p} - m
-\not{b}\ga^5](1 +\al
 I_{\infty}) -3m I_{\infty}\, -3
\not{c}\,\ga^5I_{\infty}\ + }\\ & + & {\rm regular\,(}\mu^2\ {\rm
independent)\,terms}, \quad
 {\rm where}\ \ I_{\infty}  =  \frac{\hbar e^2}{16\pi^2}[\frac{2}{4-D} - ( C
+
\log\frac{m^2}{4\pi\mu^2})]\nnb \,;
\eeqa
 $C$ is the  Euler  constant and $\mu^2$ the scale needed in the dimensional
scheme,
$$\dst
\frac{d^4k}{(2\pi)^4} \Rightarrow
\mu^{(4-D)}\frac{d^Dk}{(2\pi)^D}\,.$$

\noindent In such a calculation \footnote{\,We used
$\ga_{\mu}\ga^{\mu} = D$ and $k_{\mu}k_{\nu} = k^2\,g_{\mu\nu}/D\,:$ this
allows shifting of internal momenta and ensures gauge invariance. Other
prescription would lead to different finite counterterms (spurious
anomalies). This
point is sometimes missed, in particular in reference \cite{ChungOh} ( after
eq. 7), leading to differences proportional to $4-D$ which give extra finite
contributions, when involved into divergent integrals.}, as there is no need
for traces of Dirac matrices, there is no unconsistency in using a fully
anticommuting
$\ga^5$ matrix.  So we have the standard Q.E.D. renormalisations of the
fermionic field (wave function and mass) and a $c^{\mu}$ dependent
renormalisation of the
$b^{\mu}$ parameter in ${\cal{L}}_1\,.$

Higher orders in
$b^{\mu}$ and
$c^{\nu}$ are given by convergent integrals, then
contribute as regular ($\mu^2$ independant) functions at D = 4, and do not
change the result (\ref{propfermion}). Of course, should one implement the
normalisation condition (\ref{nor1}), extra finite counterterms renormalizing
${\cal{L}}_1$ should be required. Indeed, Lorentz covariance and
(\ref{propfermion}) ensures that
\beqa\label{norm2} -
\frac{i}{4} Tr\left[\ga^{\mu}\ga^5<\psi(p)\bar{\psi}(-p)>^{prop.}
\right]\mid_{p=0}^{{\cal{O}}(\hbar)}
& \simeq & (\al b^{\mu} + 3c^{\mu})I_{\infty} + \\ & + &
b^{\mu}a_1[\frac{b^2}{m^2},\,\frac{c^2}{m^2},\,\frac{b.c}{m^2}] +
c^{\mu}a_2[\frac{b^2}{m^2},\,\frac{c^2}{m^2},\,\frac{b.c}{m^2}]\,.\nnb\eeqa

\noindent Note also that, as we choose to fix the wave function and the mass
of the electron by the usual normalisation conditions { at vanishing
$b^{\mu}$ and
$c^{\nu}\,,$} no finite new counterterms, such as
$b^{\al}b^{\be}[\bar{\psi}\ga_{\al}\partial_{\be}\psi]/m^2\,,$ will be
required. At this point, additive renormalisation requires a
$\hbar\De{\cal{L}}_1(c,\,b)$ counter-lagrangian.

\subsubsection{The one-loop photon self-energy}

In the same way, the self energy of the photon may be expanded in powers of
$b^{\mu}$  (up to one-loop order, ${\cal L}_2$ does not modify the photon
self energy, except at the tree level). The
$b^{\mu}$ independant contribution comes from ordinary QED, the
$b^{\mu}$ linear one being given by the two ``triangle-like" graphs with a
zero-momentum ``axial vertex"~:
\beqa\label{triangles}
 I_{\mu\nu\al}(p) & = & -\hbar e^2
b^{\al}\int\frac{d^4q}{(2\pi)^4}\frac{Tr[\ga_{\mu}(\not{q}+m)\ga_{\nu}
(\not{q}+\not{p}+m)\ga_{\al}\ga^5(\not{q}+\not{p}+m)]}{(q^2 - m^2)[(q+p)^2 -
m^2]^2}\,,\nnb \\{\rm and}\quad & - & \hbar e^2
b^{\al}\int\frac{d^4q}{(2\pi)^4}\frac{Tr[\ga_{\mu}(\not{q}+m)\ga_{\al}\ga^5
(\not{q}+m)
\ga_{\nu} (\not{q}+\not{p}+m)]}{[q^2 - m^2]^2((q+p)^2 - m^2)} = \\ & = &
-\hbar e^2 b^{\al}\int\frac{d^4q}{(2\pi)^4}\frac{Tr[\ga_{\nu}
(\not{q}+\not{p}+m)\ga_{\mu}(\not{q}+m)\ga_{\al}\ga^5(\not{q}+m)]}{[q^2 -
m^2]^2((q+p)^2 - m^2)} \stackrel{(q \rightarrow q-p)}{=}
I_{\nu\mu\al}(-p)\,,\nnb \eeqa

\noindent as soon as shift of loop momenta is allowed and thanks to the
cyclicity of the trace of a product of matrices (this last property is
independant of the dimension of the matrices and should hold in any
consistent dimensional regularisation).
\noindent The only algebraic properties we need for the calculation in
D-dimension are the algebraically consistent ones
\cite{{DR2},{BM},{GB1990}} :
\begin{itemize}
\item as previously emphasized, $\ga_{\mu}\ga^{\mu} = D$ and
$k_{\mu}k_{\nu} = k^2\,g_{\mu\nu}/D\,,$

\item the trace of $\ga^5$ times an odd number of Dirac matrices vanishes,
\item  the trace of $\ga^5$ times an even number of Dirac matrices can be
reduced with the Clifford algebra to the quantity
$Tr[\ga_{\mu}\ga_{\nu}\ga_{\al}\ga_{\be}\ga^5]\,;$ consistency enforces
$Tr[\ga_{\al}\ga_{\be}\ga^5] = Tr[\ga^5] = 0\,.$
\end{itemize}

\noindent Using Feynman parameters to combine the denominators in
(\ref{triangles}), the first triangle contribution gives
$$-\frac{\hbar e^2}{8\pi^2}b^{\al}p^{\be}
Tr[\ga_{\mu}\ga_{\nu}\ga_{\al}\ga_{\be}\ga^5]\times(A+B)$$ where the
divergent part comes from :
$$A =
\int_0^1\,d\,x[x(4-6x)
-(4-D)x(2-x)]\int\frac{(4\pi\mu^2)^{2-D/2}d^{D}q'}{\pi^{D/2}}
\frac{q'^{\,2}/D}{[q'^2 +x(1-x)p^2 -m^2]^3}\,,$$ and $B\,,$ a convergent one,
may be evaluated at
 $D = 4\,:$
$$B = -
\int_0^1\,d\,x[x^2(1-x)^2 p^2 + x(2-x) m^2]\int\frac{d^4q'}{\pi^2}
\frac{1}{[q'^2 +x(1-x)p^2 -m^2]^3}\,.$$ As $\dst \int_0^1 d\,x\ x(4-6x) =
0\,,$ the divergent part vanishes and, after Wick rotation and  integration
on
$q'\,,$ one is left with the finite quantity :
$$A+B = \frac{i}{2}\int_0^1\,d\,x\left[x(2-3x)\log{\frac{x(1-x)p_E^2 +
m^2}{4\pi\mu^2}} - x(2-x) - \frac{ x^2(1-x)^2 p_E^2 - x(2-x)
m^2}{x(1-x)p_E^2 + m^2}\right]\,.$$ An integration by parts on $x$ for the
$\log$ term finaly gives :
$$A+B = -\frac{i}{2} p^2
\,K(p^2) = -i p_E^2\int_0^1\,d\,x\,\frac{ x^2(1-x)}{x(1-x)p_E^2 + m^2}\
\stackrel{p^2 \rightarrow 0}{\simeq} i\frac{p^2}{12m^2}\,,$$ a finite result,
which moreover vanishes for
$p^2 = 0\,.$  The complete one-loop 2-photon proper Green function is :
\beqa\label{propphoton}  <A_{\mu}(-p)A_{\nu}(p)>\mid^{prop.}_{transverse} &
= &
\nnb \\ -i\,(g_{\mu\nu}p^2 - p^{\mu}p^{\nu})[ 1 & - &
\frac{4}{3}\,I_{\infty} + {\rm regular\,(}\,\mu^2\ {\rm independant)\,terms}
\, ]
\nnb \\
 -  \epsilon_{\mu\nu\rho\si}p^{\si}[2c^{\rho} + b^{\rho}\frac{\hbar
e^2}{2\pi^2}
\, p^2
\,K(p^2)] &  , & \quad  p^2\,K(p^2)\ {\rm is\ analytic\
for\ } p^2  < 4\,m^2\,,
\\
  p^2\,K(p^2) = 1 & - & \frac{\log{[\rho + \sqrt{1 +\rho^2}]}}{\rho
\sqrt{1 +\rho^2}}
\simeq \frac{p_E^2}{6m^2}\,{\rm when}\ p_E^2
\rightarrow 0 \ ({\rm with}\ \rho = \sqrt{\frac{p_E^2}{4m^2}}).  \nnb
\eeqa Note that, here also, contributions with higher order dependence on
$b^{\mu}$ are given by convergent integrals (then contributing as $\mu^2$
independent terms), thanks to gauge invariance ( the unique logarithmically
divergent polynomial term
$b^{\al}b^{\be}\Pi_{\mu\nu\,,\al\be}$ cannot be transverse on $p^{\mu}$ and
$p^{\nu}).$

A few remarks are in order to understand the results :
\begin{itemize}
\item The finiteness does not come from a cancellation between the two
``triangle-like" graphs~: on the contrary, each of the triangle being
superficially linearly divergent, its divergent part is a dimension-1
polynomial in masses and external momentum $p\,.$ The only possible
structure (in any regularisation that respects Lorentz
covariance) is
$\epsilon_{\mu\nu\al\be}\,p^{\be}\,,$ then the divergence is at most a
logarithmic one ; moreover, thanks to Bose symmetry ($\mu \leftrightarrow
\nu\,,\ p \leftrightarrow -p$), it doubles itself when the two triangles are
added.
\item Then, to {\bf understand} the convergence, it is {\bf necessary} to
use gauge invariance and the Ward identity (\ref{WI3}) for the
operator
$J_{\al}^5\,.$  First, note that the
${\cal O}(b^{\al})$ term is given by the axial vertex
$<A_{\mu}(q)A_{\nu}(p)J_{\al}^5(-(p+q))>^{prop.}$ for
$q+p = 0\,.$ Second, the divergent part of this vertex is a dimension-1
polynomial in masses and external momenta $p$ and $q\,:
A\,\epsilon_{\mu\nu\al\be}\,(p-q)^{\be}\,,$ thanks to Bose symmetry,
constrained by gauge invariance (\ref{WI3}),  then it vanishes. Third, the
value of this vertex, thus given by convergent integrals, is uniquely fixed,
independently of the regularisation used, as soon as the vector current
conservation is ensured (see the nice "old" discussion in Adler's 1970
lectures
\cite{Adler}),  and of course the same occurs for the photon self energy
correction
$b^{\al}[I_{\mu\nu\al}(p) + I_{\nu\mu\al}(-p)]\,.$
\item As argued by Coleman and Glashow
\cite{ColGlas}, due to analyticity, {\sl `` from the once well known theory
of Feynman-diagram singularities "}, this Chern-Simons like contribution
will vanish at
$p+q =0,\ p^2 = 0\,.$ \hfill Q.E.D.

\end{itemize}

\noindent  The finiteness
$a\ fortiori$ holds for higher orders in
$b^{\mu}$ : the only superficial divergence could come from a
$b^{\al}b^{\be}\Pi_{\al\be\,;\,\mu\nu}(p,\,-p)\,,$ but the corresponding
dimension zero polynomial divergence has to be transverse with respect to
the photon momenta
$p^{\mu}\,,$ so it must vanish. Moreover, should one implement the
normalisation condition (\ref{nor1}), no extra finite counterterm
renormalizing
${\cal{L}}_2$ should be required. Indeed, Lorentz covariance and a
generalisation of Coleman and Glashow's argument  ensure that
$$ \frac{1}{12}\epsilon^{\mu\nu\rho\si}\frac{\partial}{\partial p^{\si}}
<A_{\nu}(p)A_{\rho}(-p)>^{prop.}\mid_{p=0}^{{\cal{O}}(\hbar)}\  =\ 0$$

\noindent (consider the proper Green
function of two photons $A^{\al}(p)$ and
$A^{\be}(q)$ with one insertion of
$b^{\mu}J^5_{\mu}(-(p+q))\,,$ computed with { the complete}  Lagrangian
density
${\cal{L}}_0 + {\cal{L}}_1 + {\cal{L}}_2\,,$ {\sl i.e.} to
all orders in
$b^{\nu}\,,$ and the corresponding Ward identity (\ref{WI3})).
Note also that, as we choose to fix the wave function of the
photon by the usual normalisation conditions { at vanishing $b^{\mu}$ and
$c^{\nu}\,,$} no finite new counterterm, such as
$b^2[F_{\mu\nu}F^{\mu\nu}]/m^2$ e.t.c., will be required.

At this point,
additive renormalisation requires no
$\hbar\De{\cal{L}}_2(c,\,b)$ counter-lagrangian.

\subsubsection{The one-loop photon-electron vertex}
 Finally, simple
power counting shows that the vertex function
$<\psi(p)\bar{\psi}(q)\,A_{\mu}(-(p+q))>^{prop.}$
 has no $b^{\al}$ or $c^{\be}$ dependent divergence : so, as we choose to
fix the electron-photon vertex by the usual normalisation condition { at
vanishing $b^{\mu}$ and
$c^{\nu}\,,$} the Q.E.D. counter-lagrangian $\hbar\De{\cal{L}}_0$ ensures
finiteness and correct normalisation conditions.

\subsubsection{Higher-loop orders}

To summarise, to one-loop order,  given the classical Lagrangian
${\cal{L}}_0 + {\cal{L}}_1 + {\cal{L}}_2\,,$ renormalisation only requires
the counterterms
$\hbar(\De{\cal{L}}_0 + \De{\cal{L}}_1)\,:$ if $b^{\mu}$
is renormalised to $b_1^{\mu}(b,\,c)\,,$ no $\hbar{\cal L}_2$ correction has
appeared. The
Chern-Simons like term is not renormalised - even if the $c^{\mu}$
parameter is rescaled by a $Z_3^{-1}$ factor, to compensate for the usual
photon field renormalisation $\sqrt{Z_3}\,.$ Does this stay in higher orders
?

Let us suppose that, up to N-loop order, renormalisation has been
done with counterterms $\De{\cal L}_0 + \De{\cal L}_1$ and that
the only dependence on
$b^{\nu}$ and $c^{\nu}$ of the Lagrangian plus counterterms  is through the
classical ${\cal L}_2$ term and
$b_{ N}^{\mu}(b,\,c)[\bar{\psi}\ga_{\mu}\ga^5\psi](x)\,,\
b_{N}^{\mu}(b,\,c)$ being an order-N polynomial in $\hbar\,.$
\begin{itemize}
\item One of the primitively divergent Green functions is the
electron-photon vertex. As previously argued, its divergence and
normalisation condition are the same as for ordinary Q.E.D., and so taken
into account by the counterterm $\De{\cal L}_0\,.$
\item To (N+1)-order, the
$b^{\mu},\,c^{\nu}$ dependent part of the divergence of the self-energy of
the electron ( subdivergences
being properly subtracted) is proportional to
$b^{\mu}\ga_{\mu}\ga^5$ and $c^{\mu}\ga_{\mu}\ga^5\,.$ Moreover, as
\beqa\label{norm3}
-\frac{i}{4}Tr[\ga^{\mu}\ga^5<\psi(p)\bar{\psi}(-p)>^{prop.}]\mid^{{\cal
O}(\hbar^{{ N + 1}})}_{p\,=\,0}& = & b^{\mu}a_{N+1}[\frac{\mu^2}{m^2}] +
c^{\mu}b_{N+1}[\frac{\mu^2}{m^2}] + \nnb\\
+\ b^{\mu}\al_{N+1}[\frac{b^2}{m^2},\frac{c^2}{m^2},\frac{b.c}{m^2}]& +
& c^{\mu}\be_{N+1}[\frac{b^2}{m^2},\frac{c^2}{m^2},\frac{b.c}{m^2}]\,,
\eeqa
the normalisation conditions (\ref{nor1}) require only a ${\cal L}_1$ like
counterterm.
\item On the one hand, the derivative with respect to
$b^{\al}$ of the regularised photon self-energy at (N+1)-loop order is given
\cite{{Lowenstein},{Becchi}}
by :
$$ \frac{\partial}{\partial b^{\al}}\Ga_{\mu\nu}^{(N+1)}(p,-p) = - i
\left[\frac{\partial b_{ N}^{\be}}{\partial
b^{\al}}\,<A_{\mu}(p)A_{\nu}(-p)[\int
d^4x\,\{\bar{\psi}\ga_{\be}\ga^5\psi\}(x)]>^{prop.}\right]^{(N+1)} = $$
$$ = -i \sum_{l = 0}^{l= N} \left[\frac{\partial b_{ N}^{\be}}{\partial
b^{\al}}\right]^{(l)} \left[<A_{\mu}(p)A_{\nu}(-p)[\int
d^4x\,\{\bar{\psi}\ga_{\be}\ga^5\psi\}(x)]>^{prop.}\right]^{(N+1-l)}\,.$$
\begin{itemize}
\item First, this quantity is finite : the proof is the same as the one given in
Subsection (3.1.2), based on the vanishing of a dimension-one, Lorentz
covariant 3-tensor
$a_{\mu\nu\be}\,,$ polynomial in masses, parameter $b^{\al}\,,$
external momenta
$p$ and $q\,,$ transverse with respect to $p^{\mu}$ and $q^{\nu}\,,$ and
associated to the divergence of any n-loop order $(n
\le N+1)$ axial (unintegrated) vertex
$<A_{\mu}(p)A_{\nu}(q)[\bar{\psi}\ga_{\be}\ga^5\psi](-(p+q))>^{prop.}\,,$
with subdivergences properly subtracted. \item Second, we look
for possible  finite counterterms required by the normalisation conditions
(\ref{nor1}) ; following Coleman and Glashow's argument
\footnote{\,The existence of an anomaly for the axial current conservation
law does not enter the game as we used only vector current conservation.},
analyticity and gauge invariance of the n-loop order axial (unintegrated)
vertex, {\sl i.e.} vector current conservation,
 enforce its proportionality to  some
$p^{a}q^{b}\,G^{(n)}_{ab\,;\,\mu\nu\al}(p,q,b,m)\,,\ \forall n\,,$ giving
after integration
$-p^{a}p^{b}\,G^{(n)}_{ab\,;\,\mu\nu\al}(p,-p,b,m)\,.$ \end{itemize}

On the other hand, the derivative with respect to
$c^{\al}$ of the regularised photon self-energy at (N+1)-loop order is given
\cite{{Lowenstein},{Becchi}}
by :
$$ \frac{\partial}{\partial c^{\al}}\Ga_{\mu\nu}^{(N+1)}(p,-p)
 = - i\left[<A_{\mu}(p)A_{\nu}(-p)[\int
d^4x\,\{\frac{1}{2}\epsilon_{\al\rho\si\la}
F^{\rho\si}A^{\la}\}(x)]>^{prop.}\right]^{N+1} $$
$$ -i \sum_{l = 0}^{l= N}
\left[\frac{\partial b_{ N}^{\be}}{\partial c^{\al}}\right]^{(l)}
\left[<A_{\mu}(p)A_{\nu}(-p)[\int
d^4x\,\{\frac{1}{2}\epsilon_{\al\rho\si\la}
F^{\rho\si}A^{\la}\}(x)]>^{prop.}\right]^{(N+1-l)}\,.$$
In that equation, the vertex insertions are at least at one loop order, so
the same argument as before holds, thanks to the Ward identity
(\ref{WI3bis}).

 As a consequence, to
(N+1)-loop order, the quantity
$\dst\frac{1}{12}\epsilon^{\mu\nu\rho\si}\frac{\partial}{\partial p^{\si}}
<A_{\nu}(p)A_{\rho}(-p)>^{prop.}\mid_{p=0}$ is
given by its value for vanishing
$b^{\al}$ and $c^{\be}\,.$
Then the normalisation conditions (\ref{nor1}) may be ensured without any
$\De{\cal L}_2$ counterterm.

\item  The $b^{\mu}$ and  $c^{\nu}$ independent part of the electron and
photon self-energy
 being correctly renormalised by the counterterm $\De{\cal
L}_0\,,\
\De{\cal L}_0 + \De{\cal L}_1$ ensures the correct renormalisation up to
order (N+1).  \hfill{Q.E.D.}
\end{itemize}
\noindent So, to all orders of perturbation theory, the
theory described by ${\cal L}_0 + {\cal L}_1 + {\cal L}_2$ is a quantum
consistent theory, with an (infinite) renormalisation of the photon field,
electron mass, electric charge and $b^{\mu}$ parameter, and {\bf no ${\cal
L}_2$ renormalisation.}

\subsection{Gauge invariant QED with a CPT-odd, Lorentz breaking term.}

Consider now a possible situation without ${\cal L}_2$ in the  classical
lagrangian, $i.e.$ with a value zero for the parameter
$c^{\mu}$ defined by the normalisation condition (\ref{nor1}). The
analysis of the previous Subsection shows that ${\cal L}_2$ is not
renormalised : then, its absence at the classical level is stable against
perturbative expansion, to all-loop order, and the
theory described by ${\cal L}_0 + {\cal L}_1$ is a quantum consistent
theory, with an (infinite) renormalisation of the photon field, electron
mass, electric charge and $b^{\mu}$ parameter.

\section{Discussion and summary}

We have exemplified the fact that, as soon as a theory is correctly defined
(not only by a Lagrangian density such as ${\cal L}_0\ +{\cal L}_1 + {\cal
L}_2\,,$ but by some symmetry requirement such as gauge invariance of the
action and appropriate normalisation conditions (\ref{nor1}), the quantum
corrections are unambiguous.

\noindent The opposite conclusion often given in the literature
\cite{{JK99},{Chung-Perez}}, results from an unsufficient definition of the
model and some unprecised arguments :
\begin{itemize}
\item Jackiw and  Kostelecky \cite{JK99} never introduce any regulator. Then
some of their relations are delicate ones : see for example for a divergent
integral ( after equ.12), the commutation of a derivation with respect to
external momentum and the integration. If the integral in their equation
(11), which is twice our tensor
$I_{\mu\nu\al}(p)\,,$ is computed with dimensional regularisation, a result
$\dst\left[\frac{\theta}{\sin{\theta}} -1\right]$ is found, and not simply
$\dst\frac{\theta}{\sin{\theta}}$ (with
$p^2 = 4 m^2\sin^2{\theta/2}\,).$

\item Moreover, in the absence of normalisation conditions or
Ward identities fixing some ambiguities, the difference of two equivalent
linearly divergent integrals gives an ambiguous logarithmic divergent one.
Even when one uses a symmetric integration that suppresses the linearly (and
eventually the logarithmicaly) divergent part, the finite part remains
ambiguous. The ``surface term" that comes from a shift in the integration
momentum in a linearly divergent integral is a regulator dependent quantity
:  if one redoes the calculation in the appendix A5-2 of Jauch and Rohrlich
standard book
\cite{JR} with the dimensional scheme, one easily checks that no "surface
term" occurs after a shift of the integration momenta
\cite{DR2}). Recall that this possibility of shifting internal momenta is
needed to preserve gauge invariance in loop calculations (see for example
\cite[subsect.17.9]{BD2}).

\item If the gauge invariance constraint (\ref{WI3}) is not
imposed, {\bf the finiteness of the ``triangles" appears as purely
accidental}
\footnote{\,The vector current conservation (\ref{WI3a}) only imposes the
transversality of the (divergent part of the) photon self energy, which does
not excludes a
$\epsilon_{\mu\nu\al\be}b^{\al}p^{\be}$ infinite contribution.}
\cite{{JK99},{Chung-Perez}} {\bf and would not stay in higher orders}. So,
some authors rightly conclude that the corresponding one-loop finite
contribution is ambiguous
\cite{Chung-Perez}. Note that in the main text of
\cite{Jackiw}, Jackiw correctly summarizes the discussion of his Section 4 by
the sentence ``{\sl An arbitrary value persists only when no symmetry is
enforced}", but he gave his paper a misleading title :``{\sl When radiative
corrections are finite but undetermined}". Such unexpected finiteness also occurs in other situations, for exemple in the so-called complex sine-Gordon model : as was shown in \cite{bonneaudelduc} the origin lies, not in  some isometry of the theory, but in the physical property of non-production for the S-matrix.

\item Although local gauge invariance is lost at the level of the Lagrangian
density, the breaking is linear in the quantum fields, and then the
invariance of the action ensures the validity of the usual Ward identity that
corresponds to vector current conservation. So, the difference advocated by
Jackiw and others between a term in the Lagrangian density locally gauge
invariant and one giving a gauge invariant contribution to the action is not
relevant for the present case, as, but for the tree level, such Green
functions with insertion satisfy the same usual Ward identity (\ref{WI4}).

\item Finally, it is difficult to see the difference advocated in \cite{JK99}
between a first order (in
$b^{\mu})$ perturbative calculation and what is claimed to be a
``non-perturbative unambiguous value", but is, as a matter of fact, obtained
with exactly the same triangle integrals as everyone. More precisely, after
the expression given in \cite[equation (5)]{JK99} for a complete
(non-perturbative) contribution of the breaking to the 2-photon one-loop
Green function, it is argued that, as the linear divergences cancel among
the two terms of the integral, there will be no ``momentum routing"
ambiguity, and then a unique value will be obtained, for example by  an
expansion in the parameter
$b^{\mu}\,.$ As we previously explain, this argument is uncorrect.  Moreover, in
\cite{Chung-Perez} the computation is also done to all orders in the breaking
parameter $b^{\mu}$ and it is explicitely verified that higher orders do not
contribute to a possible correction to $c^{\mu}\,.$
\end{itemize}
Note also that the Lagrangian density
${\cal L}_0\ + {\cal L}_2$ would not lead to a coherent theory as an
(infinite) counterterm ${\cal L}_1$ appears at the one-loop order
(\ref{propfermion}).

To summarize, we have proven that a theory with a vanishing tree
level Chern-Simons like breaking term is consistent as soon as it is
correctly defined : thanks to the gauge invariance of the action, the
normalisation condition
$c^{\mu} = 0$ may be enforced to all orders of perturbation theory. A ${\cal
L}_2$ term, bilinear in the gauge field, appears in facts as a minor
modification of the {\bf gauge fixing term} as $\partial_{\nu}A^{\nu}$
remains a free field : then, as part of the ``gauge term", it is,  as usual,
not renormalised .

\vspace{5mm}
Of course, the 2-photon Green function receives {
definite radiative corrections} $\quad\dst\simeq \frac{\hbar e^2}{12\pi^2}
\frac{p^2}{m^2}\epsilon_{\mu\nu\rho\si} \,p^{\si}b^{\rho} + \cdots $ Recall
the case of the electric charge : physically measurable quantities occur only
through the
$p^2$ dependence of the photon self-energy (as the Lamb-shift is a
measurable consequence of a non-measurable charge renormalisation).
Unfortunately, as Coleman and Glashow explained, the absence of
birefringence of light in vacuum,
$i.e.$ the vanishing of the parameter $c^{\nu}\,,$ gives no
constraint on the value of the other one $b^{\mu}\,.$

\bibliographystyle{plain}

\end{document}